\documentclass[graybox]{svmult}
\usepackage{mathptmx}       
\usepackage{helvet}         
\usepackage{courier}        
\usepackage{type1cm}        
\usepackage{makeidx}         
\usepackage{graphicx}        
\usepackage{multicol}        
\usepackage[bottom]{footmisc}
\usepackage{url}

\usepackage{array}

\makeindex             

\begin{document}
	
	\title*{A Percolation-based Thresholding Method with Applications in Functional Connectivity Analysis}
	\titlerunning{Percolation-based Thresholding Method} 
	\author{Farnaz Zamani Esfahlani \and Hiroki Sayama}
	\institute{Farnaz Zamani Esfahlani \at Center for Collective Dynamics of Complex Systems, Department of Systems Science and Industrial Engineering, Binghamton University, P.O. Box 6000, Binghamton, NY 13902-6000, USA, \email{fzamani1@binghamton.edu}
	\and Hiroki Sayama \at Center for Collective Dynamics of Complex Systems, Department of Systems Science and Industrial Engineering, Binghamton University, P.O. Box 6000, Binghamton, NY 13902-6000, USA, \email{sayama@binghamton.edu}}
	%
	%
	\maketitle
\abstract*{Despite the recent advances in developing more effective thresholding methods to convert weighted networks to unweighted counterparts, there are still several limitations that need to be addressed. One such limitation is the inability of the most existing thresholding methods to take into account the topological properties of the original weighted networks during the binarization process, which could ultimately result in unweighted networks that have drastically different topological properties than the original weighted networks. In this study, we propose a new thresholding method based on the percolation theory to address this limitation. The performance of the proposed method was validated and compared to the existing thresholding methods using simulated and real-world functional connectivity networks in the brain. Comparison of macroscopic and microscopic properties of the resulted unweighted networks to the original weighted networks suggests that the proposed thresholding method can successfully maintain the topological properties of the original weighted networks. }

\abstract{Despite the recent advances in developing more effective thresholding methods to convert weighted networks to unweighted counterparts, there are still several limitations that need to be addressed. One such limitation is the inability of the most existing thresholding methods to take into account the topological properties of the original weighted networks during the binarization process, which could ultimately result in unweighted networks that have drastically different topological properties than the original weighted networks. In this study, we propose a new thresholding method based on the percolation theory to address this limitation. The performance of the proposed method was validated and compared to the existing thresholding methods using simulated and real-world functional connectivity networks in the brain. Comparison of macroscopic and microscopic properties of the resulted unweighted networks to the original weighted networks suggests that the proposed thresholding method can successfully maintain the topological properties of the original weighted networks. \keywords{Percolation, Thresholding, Weighted Networks, Functional Connectivity}}

	\section{Introduction}
	Network science has become an integral part of analyzing complex systems whose aggregate behavior cannot be explained by the summation of their parts \cite{barabasi2013network}. This is in part due to the rapid advancement of data acquisition techniques that enable empirical measurement from components of complex systems, where these measurements can be used to estimate the links (similarities) between the system components. For example, neuroimaging data such as functional magnetic resonance imaging (fMRI) and electroencephalogram (EEG) are extensively used in computational neuroscience to study the connectivity between different regions of the brain \cite{rubinov2010complex}. However, since the empirical data often include measurement noise and the connection between components of the system (nodes of the network) are generally estimated using the statistical measurements, the resulted networks are dense graphs including many weak links. Analyzing such dense networks is often challenging due to the larger memory requirement, higher computational time complexity, and the limited number of measures to characterize the topology and properties of the weighted networks. Hence, usually, the weighted networks are mapped to an unweighted counterpart by binarizing the edge weights using a specific threshold where the connection weights smaller than the predefined threshold are discarded.
	
	Various thresholding methods have been introduced in the literature, which can be categorized into two main categories of ``absolute thresholds" and ``proportional thresholds" \cite{van2017proportional}. The absolute thresholding methods use a fixed threshold value to binarize the weighted networks, whereas the proportional methods generally use some statistics of the connection weights (such as mean, median, or the $p$-th percentile) for thresholding the weighted networks. However, none of these methods take into account the topological integrity of the original weighted network, which becomes problematic when some key edges that are critical for maintaining the macroscopic and microscopic properties of the network are removed. This is especially important as it has been shown that such topological changes can significantly impact the derived network metrics such as centrality measures \cite{van2010comparing}. Furthermore, the majority of thresholding methods in the literature neglect the importance of weak links, which has been shown to provide useful information about the underlying properties of the network \cite{alexander2010disrupted,granovetter1983strength}. 
	
	According to our best knowledge, the only method that has partially addressed the previous challenges is the 99\% connectedness method by Bassett et al. \cite{bassett2006adaptive}. Even though this method takes into account the overall connectedness of the nodes, it assumes the original weighted network is dense. In other words, for sparser networks where the average node degree is very small, this method does not guarantee the topological integrity of the network. This might not be a problem for functional connectivity analysis, but it is problematic in effective connectivity analysis where the resulted weighted networks are generally sparse. 
	
	To tackle these issues, here we propose a new thresholding method based on the percolation theory which takes into account the whole network topology during the binarization process. More specifically, we use the maximum threshold that maintains the topological integrity of the original network, which in this case is defined as the size of the largest connected component in the network. The rationale behind this method is to retain a minimum number of edges that keep the network in the same level of global connectedness. The performance of the proposed method was compared to existing statistical thresholding methods using both simulated (based on a simple linear model) and real-world (Attention Deficit Hyperactivity Disorder-200 (ADHD-200) and the Center for Biomedical Research Excellence (COBRE)) datasets. According to the results, the proposed percolation-based thresholding method was capable of maintaining the topological properties of the original weighted network at both macroscopic and microscopic levels.
	
	\section{Materials and Methods}
	\subsection{Percolation-based Thresholding Method}
\begin{figure}[]
		\centering
		\includegraphics[width=0.85\columnwidth]{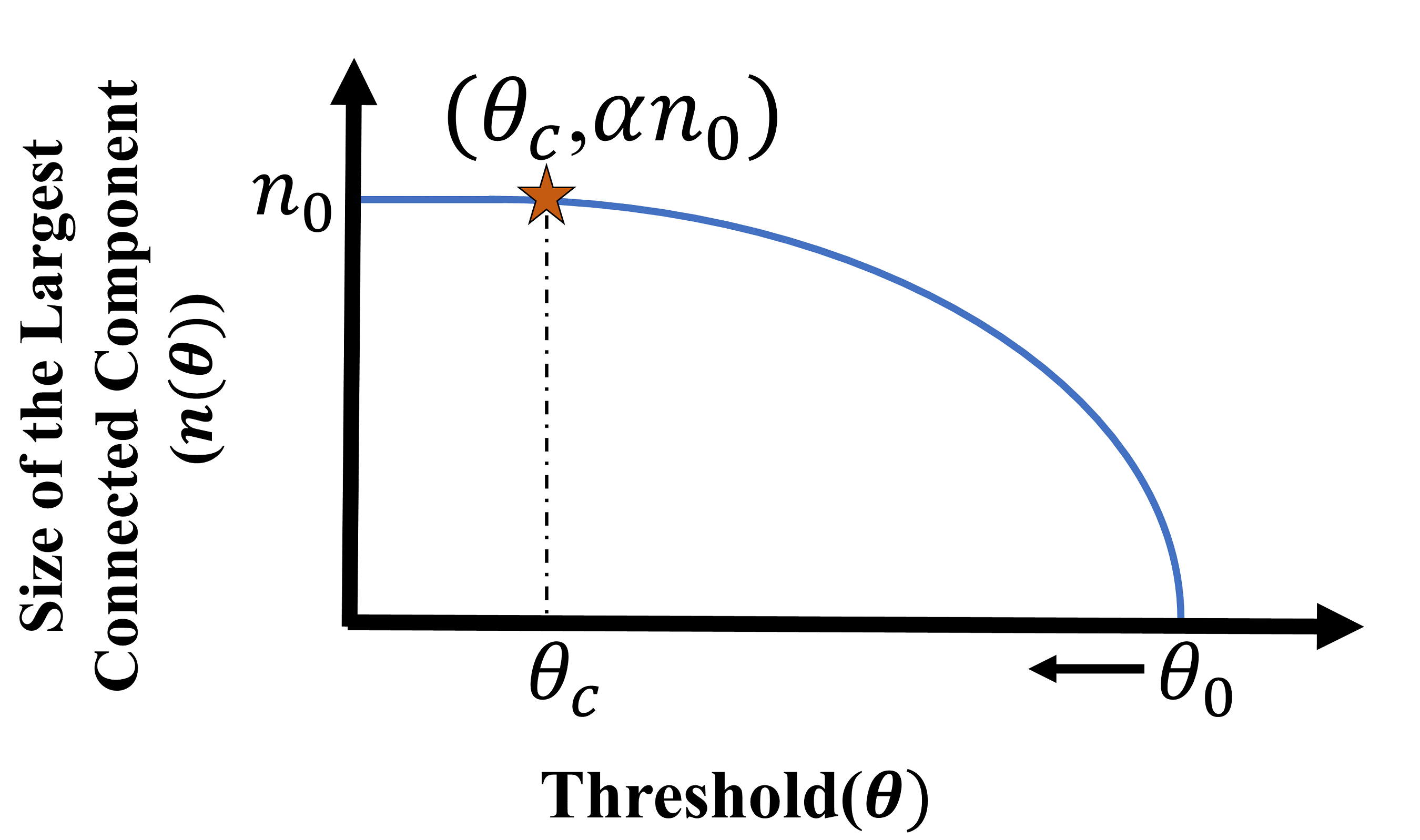}
		\caption{A schematic illustration of the percolation-based thresholding method. First, the weighted network is binarized using an initial edge weight threshold ($\theta_{0}$). Next, the largest connected component in the binarized network is characterized as a function of edge weight threshold $\theta$ and compared to the size of the largest connected component in the original network ($n_{0}$). The threshold $\theta$ is gradually decreased until the stopping criterion 
			$n(\theta_c)=\alpha n_0$ is reached.}
		\label{Framework}
	\end{figure}

	The basic idea of the percolation-based thresholding method is to identify the minimum number of edges that maintain the giant component identified in the original weighted network. To achieve this, we start binarizing the network using an initial threshold (i.e., the maximum edge weight). Next, we characterize the size of the largest connected component in the binarized network as a function of edge weight threshold $\theta$ (we call this $n(\theta))$, and compare it with the size of the largest connected component in the original network (denoted by $n_0$), which is usually the same as the number of nodes for most weighted networks constructed using real-world datasets. This process is repeated by gradually decreasing $\theta$, until the critical threshold $\theta_c$ is achieved, which is the first (largest) threshold value that satisfies
	\begin{eqnarray} 
	n(\theta_c)=\alpha n_0. 
	\label{ThetaThres}
	\end{eqnarray}
	Here $\alpha$ is the level of the connectedness of the network. In this paper, we used $\alpha = 1$ so that we can guarantee the same connectedness of the network after binarization. Relaxing this criterion (using $\alpha<1$) results in more sparse networks, which could be beneficial for specific applications. However, since we want to minimize the impact of thresholding on the topological integrity of the network, we keep $\alpha$ at one in this study. Figure \ref{Framework} provides a visual summary of the proposed percolation-based thresholding method.
	
	\subsection{Performance Evaluation}
	To evaluate the effectiveness of the thresholding method in maintaining the properties of the original weighted networks, a simple distance measure ($d$) was used to calculate the difference between the original and thresholded network properties: 
	\begin{eqnarray} 
	d=|x_{t}-x_{w}|.
	\label{Dist}
	\end{eqnarray}	
	Here $x_{t}$ represents the network property of the thresholded network and $x_{w}$ represents the same network property calculated using the original weighted network. The distance value between different macroscopic and microscopic network properties from the thresholded networks using percolation-based thresholding were compared to several proportional thresholding methods including mean, $p$-th percentile of the edge weights ($p \in \{\%99, \%95, \%75, \%50 (median), \%25,  \% 5, \% 1\}$) \cite{cohen2014analyzing}, and methods proposed by Bassett et al. \cite{bassett2006adaptive} including weighted average node degree that must be maintained at a minimum number of connected nodes, the connectedness of at least 99 \% of the nodes, and a fixed thresholding method of average degree where the average degree must not be smaller than the $\ln(|nodes|)*2$. Table \ref{treshmethods} provides a summary of the methods examined in this study. 
	
	\begin{table}[h]
		\small
		\renewcommand{\arraystretch}{1.3}
		\caption{Thresholding Methods. Label \# refers to the corresponding label in the horizontal axis of Figures \ref{thresh}, \ref{sim}, and \ref{real}.}
		\begin{tabular}{p{0.5in}p{1.5in}p{2in}>{\centering\arraybackslash}p{1in}}
			\hline
			\bf Label & \bf Method & \bf Short Description & \bf Reference\\
			\hline
			\bf 1 & Proposed \newline percolation-based method&Identify the minimum number of edges that maintain the giant component in the original weighted network.&\\
			\bf 2 & 99 $\%$ connectedness &	At least 99\% of the nodes of the network must be connected.& \cite{bassett2006adaptive}\\
			\bf 3,4 & Average degree& a) Weighted average node degree that must be maintained at a minimum number of connected nodes \newline b) The average degree must not be smaller than the $\ln(|nodes|)*2.$&\cite{bassett2006adaptive}\\
			\bf 5,6,7,\newline 8,9, \newline 10, 11 & $p$-th percentile  \newline $p \in \{\%99, \%95, \%75, \newline \%50 (median), \%25, \newline \%5,  \% 1\}$&Edges less than the $p$-th percentile value are discarded.& \cite{cohen2014analyzing}\\
			\bf 12& Mean&Edges less than the mean value are discarded.& \cite{cohen2014analyzing}\\
			\hline			
		\end{tabular}
		\label{treshmethods}
	\end{table}
	
	\subsection{Datasets}
\begin{figure}[h]
		\centering
		\includegraphics[width=0.9\columnwidth]{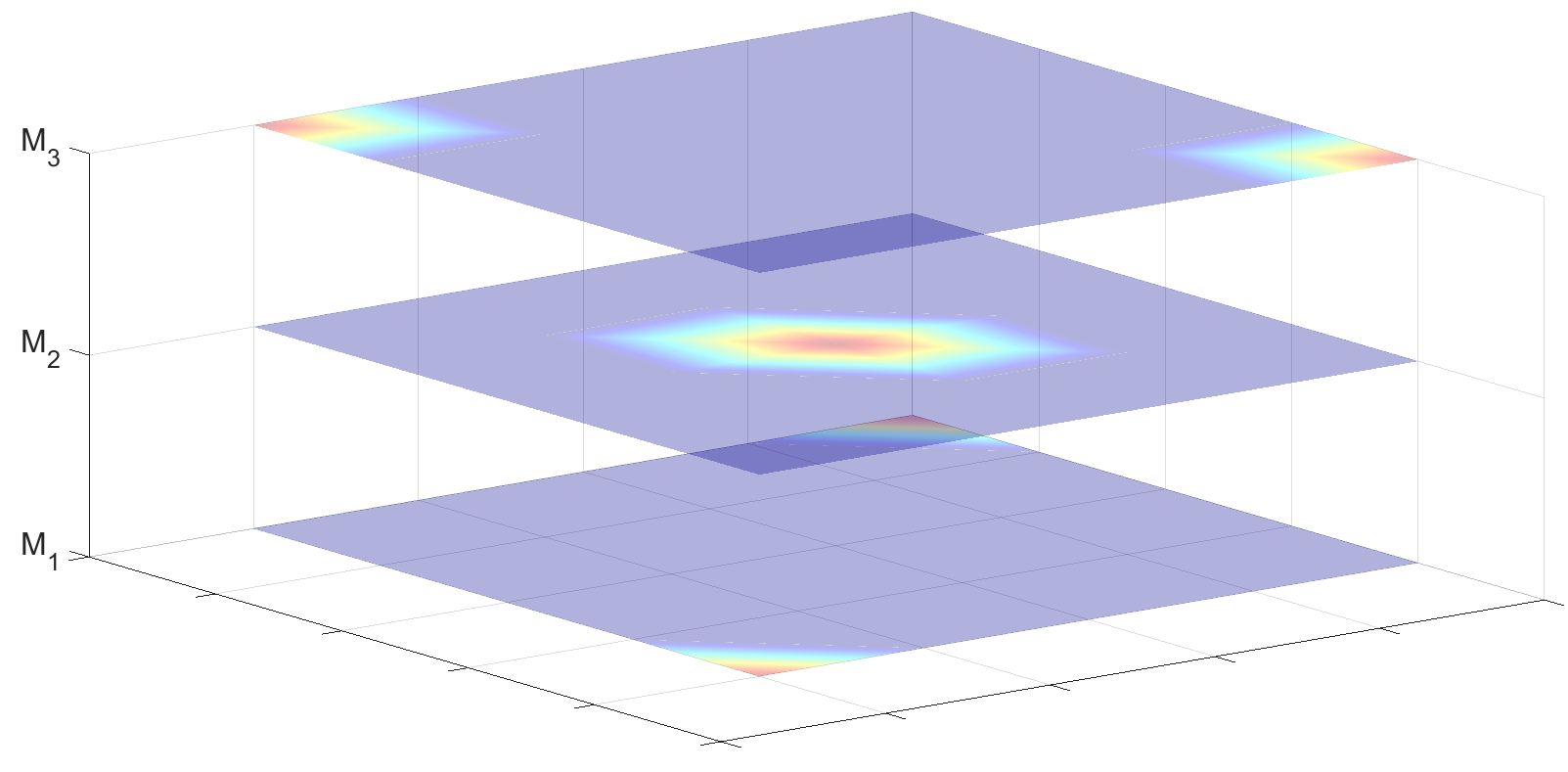}
		\caption{Simulated active brain regions.}
		\label{activbrain}
	\end{figure}

	We have used three simulated datasets and two real-world datasets to test the performance of the proposed percolation-based thresholding method. The simulated datasets with node sizes of 64, 125, and 216 (i.e., the number of Regions of Interest (ROI)) were generated using a simple linear model with random design matrices as described in \cite{michel2011total}. More specifically we have
	\begin{eqnarray} 
	y=MB+e,
	\label{simdata}
	\end{eqnarray}	
	where $B$ is the weight matrix, $M$ is the design matrix, and $e$ is the random noise. Here, $B$ corresponds to a 3D image with five blocks at the corners and one in the middle to simulate active brain regions (Figure \ref{activbrain}), $M$ is random normal variables smoothed with Gaussian fields to imitate the observed fMRI data, and $e$ is the Gaussian random noise chosen such that we have a  signal-to-noise ratio of 10 dB.
	
	For the real-world datasets, we used two major public fMRI datasets including ADHD-200 \cite{ADHD200} and the COBRE dataset \cite{COBRE}. The ADHD-200 dataset includes fMRI scans from Attention Deficit Hyperactivity Disorder (ADHD) patients and Typically Developing (TD) Children, whereas the COBRE dataset includes scans from schizophrenia patients. The ADHD-200 dataset was preprocessed according to the Athena pipeline \cite{bellec2017neuro} using the Analysis of Functional NeuroImages (AFNI) and the FMRIB's Software Library (FSL) tools, whereas the COBRE dataset was preprocessed according to the CIVET pipeline \cite{ad2006civet} using the NeuroImaging Analysis Kit (NIAK). For each dataset, the weight of edges between the network nodes (in this case the number of ROI) was estimated using correlation, partial correlation, and tangent connectivity measures \cite{craddock2013imaging}. The summary of networks used in this study is shown in Table \ref{dataset}.
	
	\begin{table}[t]
		\centering
		\small
		\renewcommand{\arraystretch}{1.3}
		\caption{Datasets.}
		\begin{tabular}{p{1.8in}>{\centering\arraybackslash}p{1.3in}>{\centering\arraybackslash}p{.8in}>{\centering\arraybackslash}p{.7in}}
			\hline
			\bf Data & \bf Sample Size \newline  (\# of Networks) & \bf Total \newline \# of Nodes & \bf Reference\\
			\hline
			Simulated&20&64, 125, 216 &\cite{michel2011total}\\
			Attention Deficit Hyperactivity \newline  Disorder (ADHD)&20&114&\cite{ADHD200}\\
			Schizophrenia (SZ) &72&39&\cite{COBRE}\\
			Typically Developed \newline Children (TD)&20&114&\cite{ADHD200}\\
			\hline			
		\end{tabular}	
		\label{dataset}
	\end{table}
	
	\section{Results}
\begin{figure}[]
		\includegraphics[width=1\columnwidth]{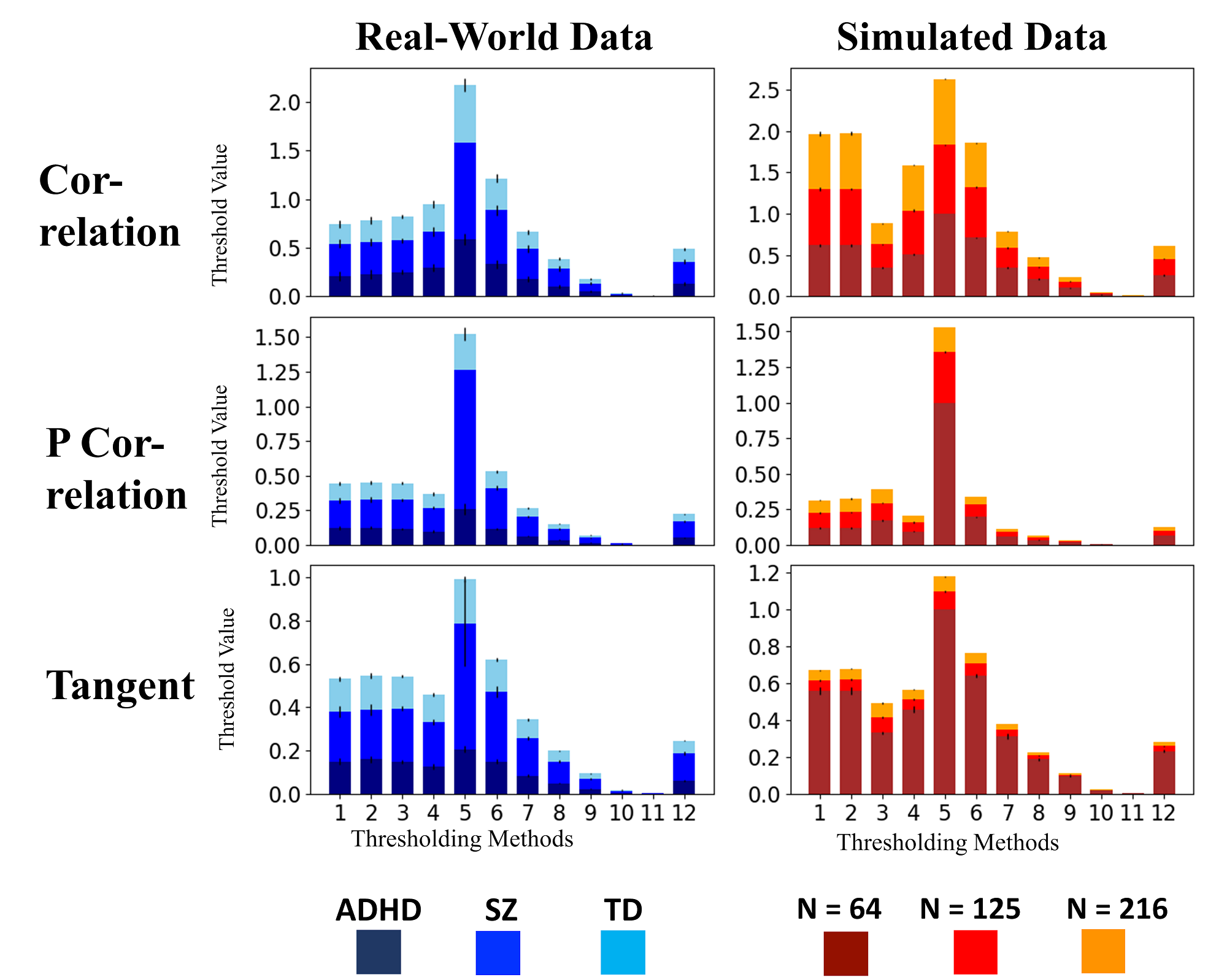}
		\caption{The threshold values obtained from thresholding methods for three real-world (ADHD, SZ and TD children) and three simulated datasets with node size $N \in \{64, 125, 216\}$. The horizontal axis in each plot shows the thresholding methods presented in Table \ref{treshmethods}. Each rows refers to a connectivity measure including: correlation which is the simplest connectivity method that quantifies the linear interdependency of two time series data, partial correlation which removes the effect of controlling random variables when calculating the interdependency \cite{wang2016efficient}, and the tangent that uses residuals of correlation matrices in the tangent space to estimate a covariance matrix \cite{varoquaux2010detection}.}
		\label{thresh}
	\end{figure}

\begin{figure}[]
		\centering
		\includegraphics[width=1\columnwidth]{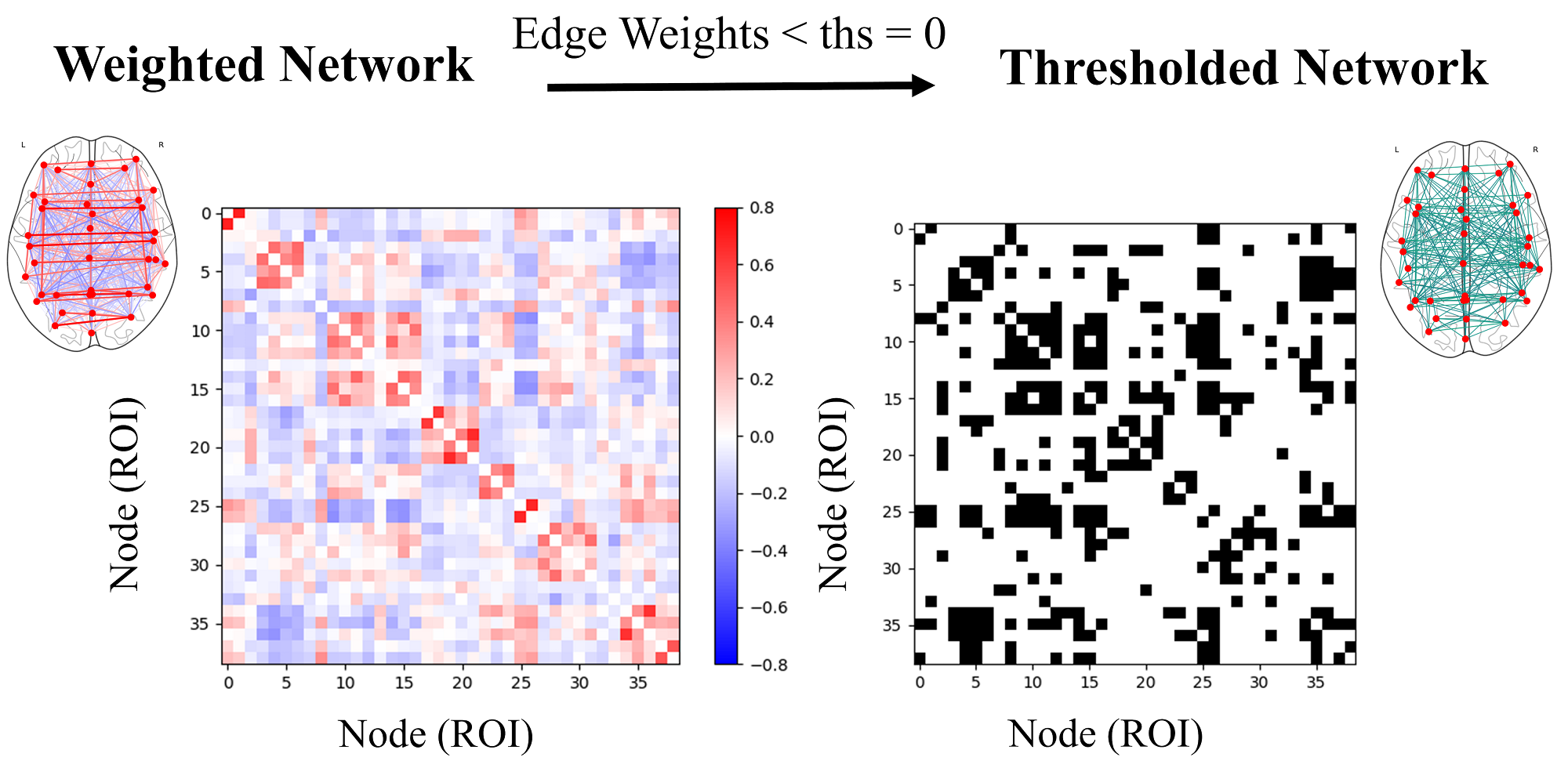}
		\caption{Example of a functional connectivity matrix/network before and after thresholding with the percolation-based thresholding method. The functional connectivity matrix represents the mean correlation for 72 networks in SZ dataset. In the weighted functional connectivity matrix, rows and columns represent ROIs (nodes), and the cell colors represent the strength of the connection between nodes. In the thresholded functional connectivity matrix, black and white cells indicate the presence and absence of the connections between ROIs, respectively.}
		
		\label{brain}
	\end{figure}

	Figure \ref{thresh} shows the mean threshold value calculated for different datasets using various thresholding methods, and Figure \ref{brain} shows an example of thresholding a weighted functional connectivity network where edges below the identified threshold value using the percolation-based thresholding method are discarded. As seen in Figure \ref{thresh}, the 99\% connectedness method provides similar threshold values to the proposed percolation-based thresholding method when using the partial correlation and tangent values, but provides a slightly higher threshold value for the correlation-based connectivity. In general, average degree based thresholding methods provided smaller threshold values except for networks obtained from the correlation measure for real-world datasets. Moreover, 99, 95 and 75 percentiles had higher threshold values, while 50, 25, 5, 1 and mean resulted in a lower threshold value than the percolation-based method. The 99\% percentile method provided the largest thresholding values, which means the binarized networks using this method will be much sparser than the binarized networks using other thresholding methods. 
	
	To understand the impact of these threshold values on maintaining the properties of weighted networks, the obtained threshold values for each weighted network was used to extract the corresponding unweighted network and several measures including the macroscopic (density, average shortest path length, and modularity) and microscopic (closeness and degree centralities) properties were calculated before and after the thresholding process. 
	
\begin{figure}[]
		\centering 
		\includegraphics[width=1.5\columnwidth,angle=90]{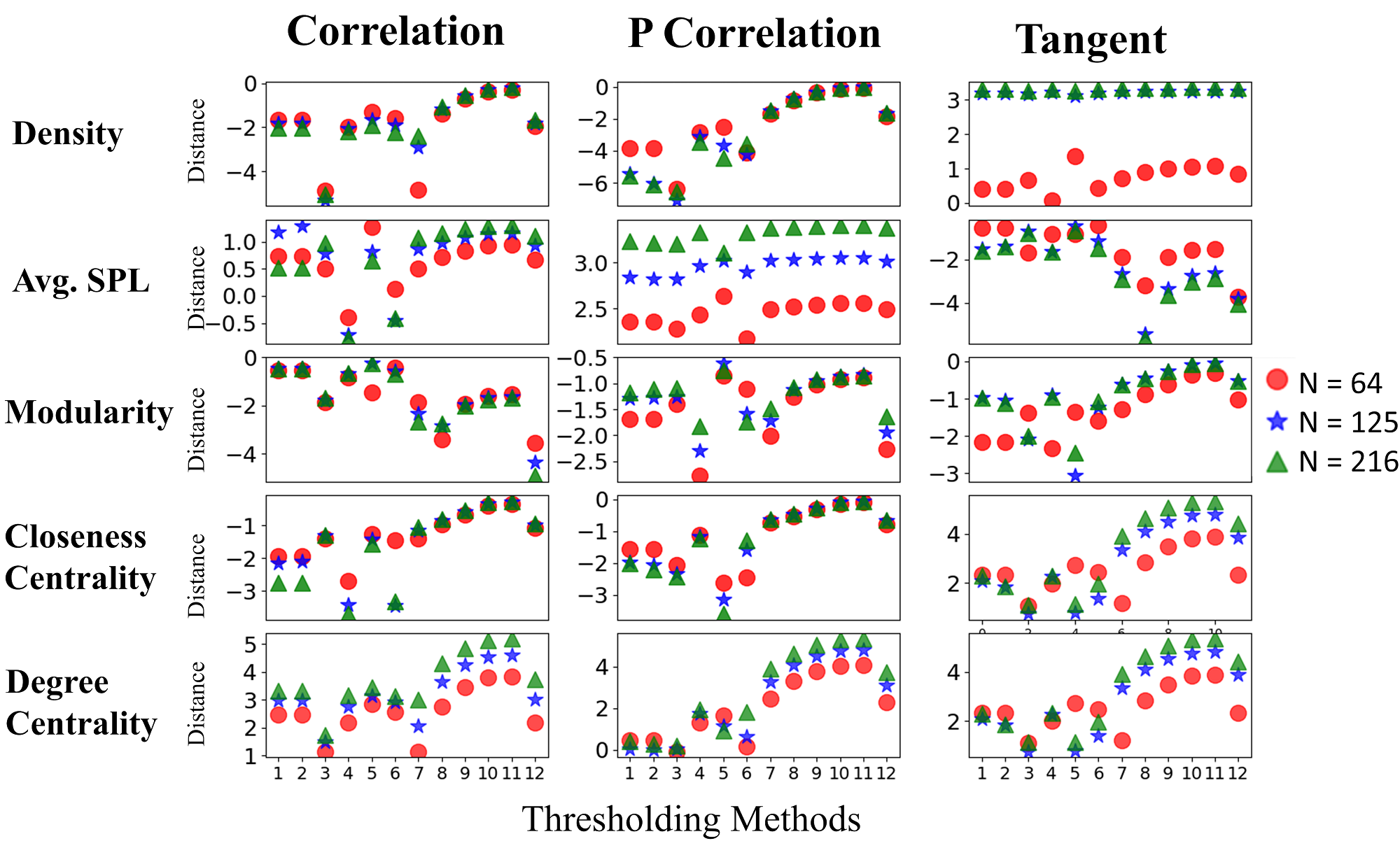}\\
		\caption{The evaluation of the macroscopic and microscopic network properties after thresholding for three simulated datasets with node size $N=[64, 125, 216]$. Each column refers to a connectivity measures (correlation, partial correlation and tangent) and each row refers to a macroscopic/microscopic network properties. In each subplot, the horizontal axis represents the different thresholding methods, and the vertical axis represents $\log$ of the distance value.} 
		\label{sim}
	\end{figure}

\begin{figure}[]
		\centering 
		\includegraphics[width=1.5\columnwidth,angle=90]{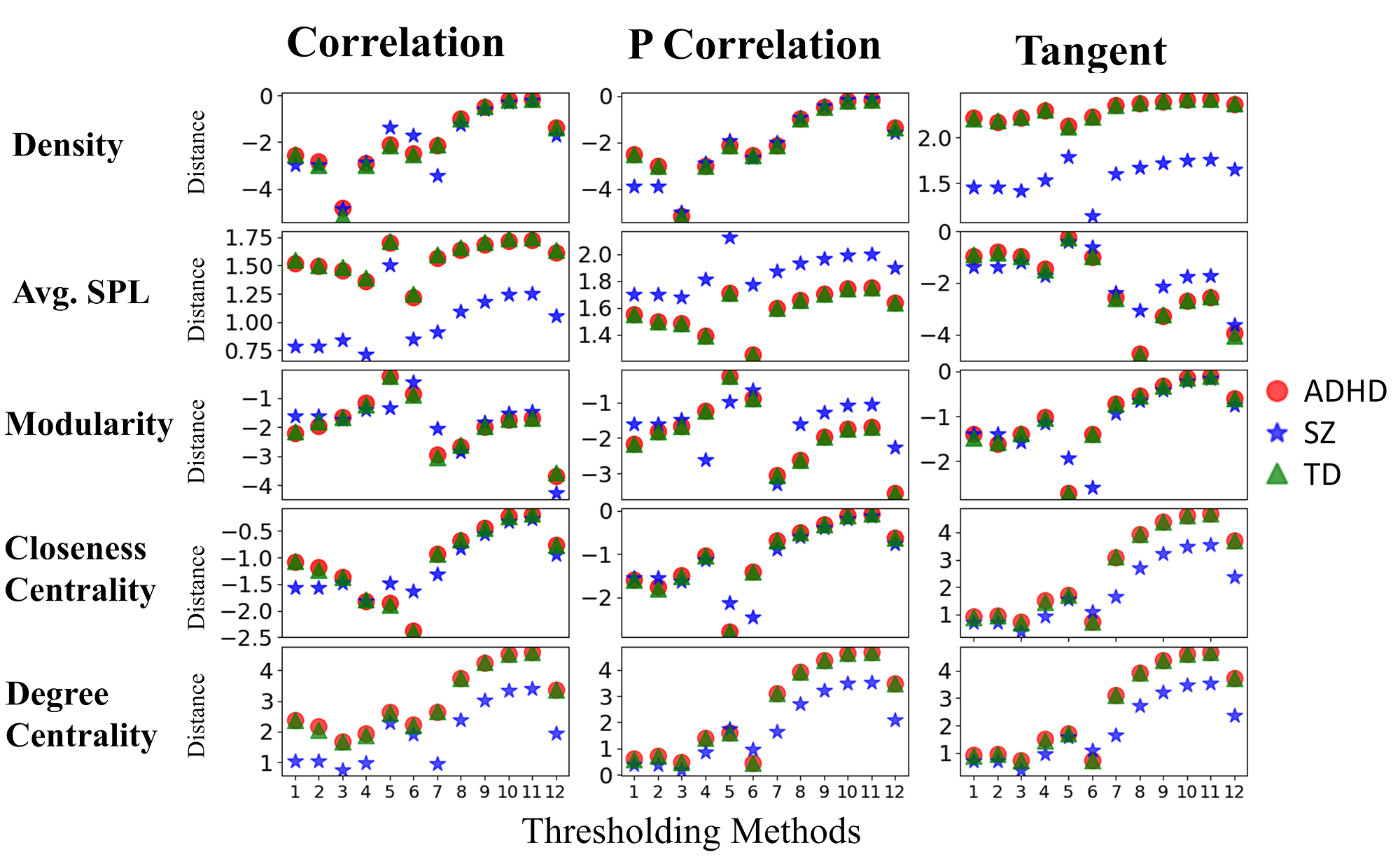}\\
		\caption{The evaluation of the macroscopic and microscopic network properties after thresholding for three real-world datasets. Each column refers to a network connectivity measure of correlation, partial correlation and tangent and each row refers to a macroscopic/microscopic network properties. In each subplot, the horizontal axis represents the different thresholding methods, and the vertical axis represents $\log$ of the distance value.} 
		\label{real}
	\end{figure}

	Figures \ref{sim} and \ref{real} show the mean distance value of the macroscopic (density, average shortest path length, and modularity) and microscopic (closeness centrality and degree centrality) network properties between the original weighted networks and the thresholded ones based on three connectivity measures for simulated and real-world datasets. According to the results, in most of the cases, the distance between the original network properties and the binarized network properties was smaller when using thresholding methods based on the network topology (percolation-based method, 99\% connectedness, and degree-based method). Having said that in some cases (e.g., mean distance of modularity calculated using tangent connectivity measure), thresholding based on $p$-th percentile (i.e., 25 \%, 5\%, and 1\%) resulted in a lower distance value between weighted and thresholded network. However, this was mainly because of selecting a very small threshold value where only small number of links were removed from the original weighted network, and hence the resulted unweighted networks were dense. Taking into account this limitation of statistical methods, the performance of the percolation-based method was especially good for preserving the modularity of the correlation and partial correlation-based networks in real-world datasets. Interestingly, the thresholding results using three tested connectivity measures showed a similar pattern in maintaining degree centrality which would indicate the robustness of degree centrality measure to outliers and noise.
	
	\section{Conclusion}
	Choosing an appropriate threshold value to convert weighted networks to unweighted ones is a major challenge in complex system analysis as the imprecise selection of such threshold could significantly alter the original network topology, which could subsequently bias the derived network properties such as centrality. In this study, we proposed a percolation-based thresholding method that maintains the topological integrity and connectivity of the original weighted network by minimizing the number of isolated nodes after binarization. More specifically, the proposed method resulted in thresholded networks with a similar macroscopic and microscopic network properties (in particular modularity, average shortest path length, and degree centrality) to the original weighted network. This is especially important in the field of network neuroscience where changes of network properties could introduce a significant bias to the outcome of the study.
	Even though the current study focuses on the neuroscience applications, the percolation-based thresholding method could also be used in other domains such as analysis of social networks and genetic networks. Examples include scientific collaboration networks where nodes are scientists and edges represent co-authorships \cite{newman2001structure}, or gene co-expression networks where nodes are genes and edges represent gene similarities \cite{spellman1998comprehensive}.
	
	The proposed method has limitations. Being an iterative procedure, the computational complexity of the proposed method is high, and it might not be suitable for very large networks. In this regard, using more efficient heuristic search methods for identifying the critical threshold value ($\theta_c$) could improve the computational time of the proposed method. Furthermore, theoretical studies that describe the relationship between different characteristics of the network and its topological integrity could be beneficial for developing more effective thresholding methods. 
\begin{acknowledgement}
	This work is supported by the National Institute of Health (NIH) grant \# MH112925-01. We would like to also thank Gregory P. Strauss and Katherine Visser for their support of this work.
\end{acknowledgement}	
	\bibliographystyle{spmpsci}
	\bibliography{mybibfile}
\end{document}